\documentclass[useAMS,usenatbib]{mn2e}

\usepackage{times}
\usepackage{natbib}
\usepackage{lscape}
\usepackage[usenames]{color}
\usepackage{graphicx}
\usepackage{amssymb}
\usepackage{pifont}
\usepackage{ulem}
\renewcommand\appendix{\par
  \setcounter{section}{0}
  \setcounter{subsection}{0}
  \setcounter{figure}{0}
  \setcounter{table}{0}
  \renewcommand\thesection{APPENDIX \Alph{section}}
  \renewcommand\thefigure{\Alph{section}\arabic{figure}}
  \renewcommand\thetable{\Alph{section}\arabic{table}}
}

\long\def\Ignore#1{\relax}

\newcommand{\kms}{\mbox{${\rm km\, s^{-1}}$}}
\newcommand{\Msun}{\mbox{$\rm M_{\odot}$}}

\newcommand{\MgI}{\mbox{$\rm Mg\,I$}}
\newcommand{\CaII}{\mbox{$\rm Ca\,II$}}

\newcommand{\sd}{\mbox{$\sigma$-drop}}

\newcommand{\eg}{{\it e.g.}}

\newcommand{\ie}{{\it i.e.}}

\title[The kinematics of \sd\ bulges from $N$-body simulations]{The
  kinematics of \sd\ bulges from spectral synthesis modelling of a hydrodynamical
  simulation }

\author[E. Portaluri et al.]
{Elisa Portaluri,$^{1}$\thanks{E-mail: elisa.portaluri@oapd.inaf.it}
Victor P. Debattista,$^{2}$ Maximillian Fabricius,$^{3,4}$ David R. Cole,$^{5}$         
\newauthor Enrico M. Corsini,$^{6,1}$, Niv Drory,$^{7}$ Andrew Rowe,$^{8}$
Lorenzo Morelli,$^{6,1}$
\newauthor Alessandro Pizzella,$^{6,1}$ and Elena Dalla Bont\`a$^{6,1}$\\
$^1$INAF--Osservatorio Astronomico di Padova, Vicolo
  dell'Osservatorio 2, I-35122 Padova, Italy\\
$^2$Jeremiah Horrocks Institute, University of Central Lancashire, 
  Preston, PR1 2HE, UK\\
$^3$Max-Planck-Institut f\"ur extraterrestrische Physik,  
  Giessenbachstra{\ss}e, D-85748 Garching, Germany\\  
$^4$Universit\"ats-Sternwarte M\"unchen, Scheinerstra{\ss}e~1,  
  D-81679 M\"unchen, Germany\\
$^5$Rudolf Peierls Centre for Theoretical Physics, Keble Road, 
  Oxford, OX1 3NP, UK\\
$^6$Dipartimento di Fisica e Astronomia `G. Galilei', Universit\`a di Padova, 
  vicolo dell'Osservatorio 3, 35122 Padova, Italy\\
$^7$Department of Astronomy, The University of Texas at Austin, 
  2515 Speedway, Stop C1400, Austin, TX 78712, USA\\
$^8$ Materials and Physics Research Centre, University of Salford, Salford, M5 4WT, UK}

\begin{document}

\date{Accepted xxx Received xxx ; in original form \today}

\maketitle

\label{firstpage}

\begin{abstract}
A minimum in stellar velocity dispersion is often
observed in the central regions of disc galaxies. To
investigate the origin of this feature, known as a \sd , we analyse the
stellar kinematics of a high-resolution $N$-body $+$ smooth particle
hydrodynamical simulation, which models the secular evolution of an
unbarred disc galaxy. We compared the intrinsic
mass-weighted kinematics to the recovered luminosity-weighted ones.
The latter were obtained by analysing synthetic spectra produced by
a new code, SYNTRA, that generates synthetic spectra
by assigning a stellar population synthesis model
to each star particle based on its age and
metallicity. The kinematics were derived from the
synthetic spectra as in real spectra to
mimic the kinematic analysis of real galaxies. We found that the
recovered luminosity-weighted kinematics in the centre of the
simulated galaxy are biased to higher rotation velocities and lower
velocity dispersions due to the presence of young stars in a
thin and kinematically cool disc, and are ultimately responsible for the \sd .

Our procedure for building mock observations and thus recovering the
luminosity-weighted kinematics of the stars in a galaxy simulation is
a powerful tool that can be applied to a variety of scientific
questions, such as multiple stellar populations in kinematically-decoupled cores
and counter-rotating components, and galaxies with both thick and thin disc components.
\end{abstract}

\begin{keywords}
  galaxies: bulges --- galaxies: evolution --- galaxies: kinematics and
  dynamics --- galaxies: nuclei --- galaxies: stellar content --- 
  galaxies: structure 
\end{keywords}

\section{Introduction}  
\label{sec:introduction}

The stellar velocity dispersion in many nearby
galaxies exhibits a central drop rather than a peak \citep{Emsellem2001,
  Comeron2008, Krajnovic2013}. Such a \sd\ was first predicted by
\citet{Binney1980} and subsequently shown to be a
common property of galaxies with a surface brightness distribution
following a S\'ersic profile \citep{Ciotti1997}.

The first \sd s were detected by \citet{Jarvis1988} in a sample of SB0
galaxies. Afterwards, this feature was detected in the late-type spiral galaxy NGC~6503 by \citet{Bottema1989},
who interpreted it as due to a central dip in the vertical velocity dispersion.
\sd s are known to be a widespread feature of the stellar kinematics in the centre of both
quiescent and active galaxies. They are observed in approximately
$30\%$ of early-type \citep{Graham1998,Krajnovic2008} and $40\%$ of
spiral galaxies \citep{Chung2004,FalconBarroso2006}. This fraction
rises to about $70\%$ for barred galaxies \citep{Perez2009,
  MendezAbreu2014}.

\citet{Koleva2008} pointed out that a metallicity mismatch of the
spectral library adopted in measuring the stellar kinematics may bias
the central velocity dispersion to lower values, producing spurious \sd s where none exist.
As a consequence, the fraction of \sd\ galaxies may be overestimated.
However, state-of-the-art methods for
measuring stellar kinematics are designed to deal with a metallicity
mismatch \citep[e.g.][]{Cappellari2004, Koleva2008, Fabricius2014} and
\sd s have been measured using a variety of techniques, spectral
libraries and wavelength ranges, suggesting that a template mismatch is not the cause of most \sd s. 

The origin and nature of \sd s are particularly important
for disc galaxies. Being related to the structure of the bulge
component and to the central mass distribution, \sd s
can be used to discriminate between the different types of bulges. Classical
bulges show centrally peaked velocity dispersion profiles,
whereas \sd s are usually observed in pseudobulges and composite
bulges, which are characterized by different components coexisting in
the same structure \citep{Fabricius2012, MendezAbreu2014,Erwin2015}. In
addition, \sd s are often associated with nuclear dust spirals and
nuclear star-forming rings \citep{Comeron2008} while their extent
does not depend on the bulge \citep{MendezAbreu2014} or galaxy size
\citep{Comeron2008}.

Different mechanisms have been proposed to explain how \sd s form and
develop. Simple dynamical models based on the epicyclic theory are not
able to explain the stellar kinematics in the central regions of
\sd\ galaxies \citep{Bottema1997, Emsellem2001}, suggesting that the
decrease in velocity dispersion is due to a cold nuclear
disc with recent star formation fueled by gas inflows.
The stars of the nuclear disc have almost the same kinematics as
the gas from which they formed, so their velocity dispersion is lower
than that of the older surrounding stars. The measured kinematics are
biased to the lower velocity dispersion of the younger stars since
they outshine the older stellar populations. Such a scenario has been
investigated by hydrodynamical simulations
\citep{Wozniak2003, Wozniak2006}. In their simulations the \sd\ forms in less than 500 Myr,
but its lifetime can exceed 1 Gyr if the nuclear region is
continuously fed with gas to sustain star formation at a level of
about 1 \Msun\ yr$^{-1}$. This easily occurs in gas-rich spiral
galaxies where a bar funnels gas towards the centre. This model explains
why \sd\ galaxies may host an inner dust system of spiral arms that
traces the path of the inflowing material.
Galaxies with a \sd\ show inner rings or nuclear discs.
The same mechanism that feeds the nuclear
structures is probably responsible for the higher fraction of Seyfert
galaxies among \sd\ galaxies \citep{Comeron2008}.
Using dissipationless simulations, \citet{Athanassoula2002} and
\citet{Bureau2005} proposed an alternative scenario where a massive
and concentrated dark matter halo removes kinetic energy from the
stellar component in the nuclear region, causing a reduction in the
velocity dispersion and the formation of a strong bar. Another
possibility is that the central \sd\ is the signature of a nuclear bar
counter-rotating with respect to the main bar of the galaxy
\citep{Friedli1996}.

The fact that \sd s are observed in unbarred galaxies leads to the conclusion that
bars may not always play a role.
\citet{deLorenzo2012} showed that a \sd\ is the signature of the
presence of a pseudobulge, which has disc properties and thus a lower
velocity dispersion with respect to the rest of the galaxy. In this
case, the \sd\ is not actually a drop with respect to the higher
velocity dispersion of the bulge, but the maximum velocity dispersion
of the pseudobulge itself.
Alternatively, the kinematics of galaxies hosting counter-rotating stellar discs is
characterized by two off-centre and symmetric peaks in the stellar
velocity dispersion in combination with zero velocity rotation
measured along the galaxy major axis. These kinematic features, which
result in a remarkably strong \sd , are observed in the radial range
where the two counter-rotating components have roughly the same
luminosity and their line-of-sight velocity distributions (LOSVDs) are
unresolved \citep{Bertola1996,Coccato2015}.
Other studies interpreted the presence of the \sd\ as due to the lack
of a central supermassive black hole or more generally of a flat
density core \citep{Dressler1990}. Other authors suggest that \sd s are
due to an underestimation of the
true velocity dispersion in galaxies hosting a central supermassive
black hole \citep{vanderMarel1994}.

 In this paper, we study the kinematic
properties of a simulated bulge characterized by a \sd , using a hydrodynamical
simulation of an unbarred spiral galaxy combined with stellar
population synthesis models. We compare the intrinsic
kinematics of the stellar component we measure directly from the
simulation to the one we recover by mimicking real integral-field
spectroscopic observations. The simulation is presented
in Section~\ref{sec:simulation} and the analysis method of the
intrinsic mass-weighted and recovered luminosity-weighted stellar
kinematics of the simulated galaxy is presented in Section
\ref{sec:analysis}. The results are discussed in
Section~\ref{sec:results} and conclusions are given in
Section~\ref{sec:conclusions}.

\section{Simulation}
\label{sec:simulation}

The simulation we use here is similar to the model presented in
\citet{Cole2014}. The model forms a disc galaxy inside a
pressure-supported gas corona within a dark matter halo.  The host
dark matter halo has a virial radius $r_{200} = 198$ kpc, virial mass
$M_{200} = 9 \times 10^{11} \Msun$ and concentration $c = 19$.  Both the
dark matter halo and the initial gas corona consist of 5 million
particles.  The dark matter particles have mass $8.5 \times 10^4
\Msun$ (those with starting radius smaller than 56 kpc, corresponding
to $90\%$ of the halo particles), while those at larger radii have a
mass $1.7 \times 10^6 \Msun$.  Dark matter particles have a softening
$\epsilon = 103$ pc.  Gas particles have softening $\epsilon = 50$ pc,
inherited by star particles forming from them; initially gas particles
all have mass $2.7 \times 10^4 \Msun$, giving the corona a mass $11\%$
that of the dark matter.  The gas corona has an initial angular
momentum such that $\lambda = 0.041$.  The initial conditions do not
include star particles, all of which form from the gas as it cools and
settles into a disc.

The simulation was evolved with the $N$-body$+$ smoothed-particle hydrodynamics
(SPH) code {\sc GASOLINE}
\citep{Wadsley2004} with an opening angle of the tree code
$\theta = 0.7$.  We use a base time step of 10 Myr with a refinement
parameter $\eta = 0.175$; gas particles also satisfy the time step
condition $\delta t_{\rm gas} = \eta_{\rm courant}h/[(1 + \alpha)c + \beta
  \mu_{\rm max}]$, where $h$ is the SPH smoothing length, $\alpha$ is the
shear coefficient, which is set to 1, the viscosity coefficient $\beta = 2$
and $\mu_{\rm max}$ is described in \citet{Wadsley2004}.
$\eta_{\rm courant} = 0.4$ is the refinement parameter for gas particles
and controls their time step size.  The SPH kernel uses the 32 nearest
neighbours.  We use the gas cooling, star formation and stellar
feedback prescriptions of \citet{Stinson2006}, which do not take
into account the effect of gas metallicity.

A gas particle is eligible to form stars if it has number density $n >
100$ cm$^{−3}$, is cooler than $T = 15,000$ K and is part of a
converging flow.  However only $10\%$ of eligible gas particles
actually form stars; these form with a mass $9.4 \times 10^3 \Msun$. A
gas particle can form multiple star particles but if its mass falls
below $21\%$ of the initial mass then its remaining mass is
distributed amongst the nearest neighbours, and the particle
deleted.

Each star particle corresponds to a stellar population with a
Miller-Scalo \citep{Miller1979} initial mass function. Stellar
feedback from Type II and Type Ia supernovae (SNeII, SNeIa) is
included, as well as from AGB stellar winds.  The effect of the
supernova feedback, which deposits $0.4 \times 10^{51}$ erg of energy
per supernova, is modelled at the subgrid level as a blast wave
through the interstellar medium \citep{Stinson2006}.  {\sc Gasoline}
tracks the production of iron and oxygen using the yields of \citet{Raiteri1996, Woosley1995} and \citet{Weidemann1987}.  Unlike
the simulation in \citet{Cole2014}, metallicity diffusion
(\eg\ \citealt{Loebman2011}) between gas particles was included in this
simulation with the diffusion parameter of \citet{Shen2010} set to $D
= 0.03$. We do not include any AGN feedback since the model does not
contain a supermassive black hole.

\section{Intrinsic mass-weighted and recovered luminosity-weighted kinematics}
\label{sec:analysis}

In this paper we generate mock observations by building spectra from
the available simulation time-steps.
We consider only the star
particles of the simulated galaxy and neglect the contribution of
gas particles to the resulting luminosity-weighted spectra, which we
analyse as if they were real, using standard data-analysis techniques.
The recovered stellar kinematics are compared with the intrinsic ones
obtained by weighting the velocity of the star particles by their
mass.

The mock spectra are generated using SYNTRA (\ie\ SYNthetic specTRA)
our code written in the {\rm PYTHON} language was designed specifically
to select stellar population synthesis
models to assign the spectral energy distribution suitable to each
star particle of the simulation time-step according to its age and
metallicity and produce final spectra with a given instrumental setup.
We use the {\sc Galaxev} library of
evolutionary stellar population synthesis models computed using the
isochrone synthesis code of \citet{Bruzual2003}. They cover the
wavelength range from 3200 \AA\ to 9500 \AA\ with a resolution of 3
\AA\ and were obtained using the evolutionary tracks by \citet{Bertelli1994}
adopting a \citet{Salpeter1955} initial mass function with a lower and
upper mass cutoff of 0.1 and 100 M$_{\odot}$,
respectively. As we do not have a
stellar population synthesis model for all the combinations of ages
and metallicities of the star particles, SYNTRA considers the bins listed
in Table~\ref{tab:bins} to assign the stellar population synthesis
model to each star particle.

The adopted stellar population models are shifted according to the
line-of-sight (LOS) velocities of the star particles and weighted by their
luminosities. We add Poissonian noise to the single mock spectra but
we did not include either read-out or background noise. Finally,
the star particles are grouped into $71\times71$ apertures of
$0.15\times0.15$ kpc$^2$ in the galaxy plane and their single mock spectra are summed,
generating a number of luminosity-weighted spectra. They are then rebinned
along the dispersion direction to a logarithmic scale with a binning
in velocity corresponding to the spectral resolution of the stellar
population models (22 \kms ). This allows SYNTRA to build a mock
integral-field spectroscopic data cube for each simulation time-step
that covers the inner $5.3\times5.3$ kpc$^2$ of the simulated galaxy for a given orientation.

\begin{table}
\caption{\small{Age and metallicity bins used for the spectral synthesis.}}  
\setlength{\tabcolsep}{1.4mm}
\label{tab:bins}
\begin{center}
\begin{small}
\begin{tabular}{cc}
\hline
\noalign{\smallskip}
\multicolumn{1}{c}{Range} &
\multicolumn{1}{c}{Adopted Value} \\
\multicolumn{1}{c}{(1)} &
\multicolumn{1}{c}{(2)} \\
\noalign{\smallskip}   
\hline
\noalign{\smallskip}
\multicolumn{2}{c}{Age  ($10^9$ Gyr)} \\
\noalign{\smallskip}
 0.025 -  0.075 &   0.050 \\ 
 0.075 -  0.150 &   0.100 \\
 0.150 -  0.300 &   0.200 \\
 0.300 -  0.500 &   0.400 \\
 0.500 -  0.700 &   0.570 \\
 0.700 -  0.900 &   0.810 \\
 0.900 -  1.200 &   1.020 \\
 1.200 -  1.600 &   1.430 \\
 1.600 -  2.250 &   2.000 \\
 2.250 -  2.750 &   2.500 \\
 2.750 -  3.500 &   3.000 \\
 3.500 -  4.500 &   4.000 \\
 4.500 -  5.500 &   5.000 \\
 5.500 -  7.000 &   6.000 \\
 7.000 -  9.000 &   8.000 \\
 9.000 - 11.000 &  10.000 \\
 11.000 - 20.000 & 13.000\\
 \noalign{\smallskip}
  \multicolumn{2}{c}{$Z$ (dex)} \\
 \noalign{\smallskip}
 0.0000 - 0.0002 & 0.0001\\
 0.0002 - 0.0020 & 0.0004\\
 0.0020 - 0.0060 & 0.0040\\
 0.0060 - 0.0140 & 0.0080\\
 0.0140 - 0.0350 & 0.0200\\
 0.0350 - 1.0000 & 0.0500\\ 
\noalign{\smallskip}       
\hline                      
\end{tabular}
\end{small}
\end{center}
\begin{minipage}{8.5cm} 
{\it Note.\/} Column (1): age and metallicity bins of the simulation star
particles. Column (2): adopted age and metallicity for the stellar population synthesis
model from \citep{Bruzual2003}.
\end{minipage}
\end{table}

\begin{figure}
\centering
\begin{tabular}{c}
  \includegraphics[width=1\hsize,angle=0]{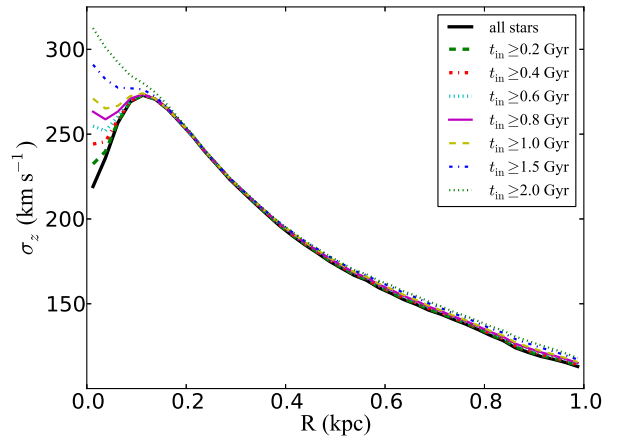}
\end{tabular}
\caption{Radial profile of the vertical velocity dispersion $\sigma_z$
  measured from the intrinsic mass-weighted kinematics as a function
  of the age of the simulation star particles. This is the azimuthal average profile
  along the $z$-direction with respect to the plane of the galaxy at 10 Gyr, as seen face-on.
  The solid black line corresponds to the profile obtained including
  all the simulation star particles while the coloured lines from
  bottom to top show to the profiles obtained without stars
  younger than $0.2, 0.4, \ldots, 1.0, 1.5$ and 2.0 Gyr as indicated
  in the top-right corner. A colour version of this figure is available online.}
\label{fig:sigmaz_age}
\end{figure}

The measurement of the stellar kinematics from spectra is
usually done by matching an observed spectral region where absorption
lines are present with a set of template spectra of either observed
stars or stellar population synthesis models at the same instrumental
spectral resolution. The relative broadening between the observed and
template spectrum then gives the galaxy velocity dispersion. 
We use the penalized pixel-fitting algorithm ({\sc pPXF},
\citealt{Cappellari2004}) with the {\sc Galaxev} library to recover
the luminosity-weighted stellar kinematics from the absorption lines
of the mock spectra in the wavelength ranges between 4750 \AA\ and 5250
\AA, which covers the \MgI\ triplet at $\lambda\ 5164, 5173, 5184$
\AA\  and between 8300 \AA\ and 8800  \AA\ covering the \CaII\ triplet at
$\lambda\ 8498, 8542, 8662$ \AA .
The {\sc pPXF} algorithm applies the maximum penalized likelihood
method to extract information from observed galaxy spectra. It works
in pixel space creating the model galaxy spectrum by convolving a
template spectrum with a Gauss-Hermite parametrized LOSVD
\citep{vanderMarel1993, Gerhard1993}.
The Gauss-Hermite moments are an effective
measure of the velocity distribution of a galaxy due to the Gaussian
nature of the first order term in the expansion of the LOSVD. The
first moment is related to the average LOS velocity, $v$, of a collection
of stars within a galaxy while the second moment is the variance of
the first moment and therefore is related to the LOS velocity
dispersion, $\sigma$. A $\chi^2$ analysis shows that this
parametrization is satisfactory for measuring the LOSVD of the
simulated galaxy and that the wide range of age and metallicity covered
by models available in the {\sc Galaxev} library at least minimizes the effect
of template mismatching in measuring the stellar kinematics
\citep[see][]{Bender1990, Koleva2008}. In this work we are interested
in the first two moments of the Gauss-Hermite parametrization.
We tested whether our results vary for the two pPXF versions available
and no significant differences were noticed when comparing $1\sigma$ errors. 

We test our measurement method by weighting the star particles'
velocity according to their mass only. The recovered mass-weighted
kinematics is very close to the intrinsic mass-weighted one, with the residuals being less than 5\%.

\section{Results}
\label{sec:results}

\begin{figure*}
\centering
\begin{tabular}{c}
 
\includegraphics[width=0.33\hsize,angle=0]{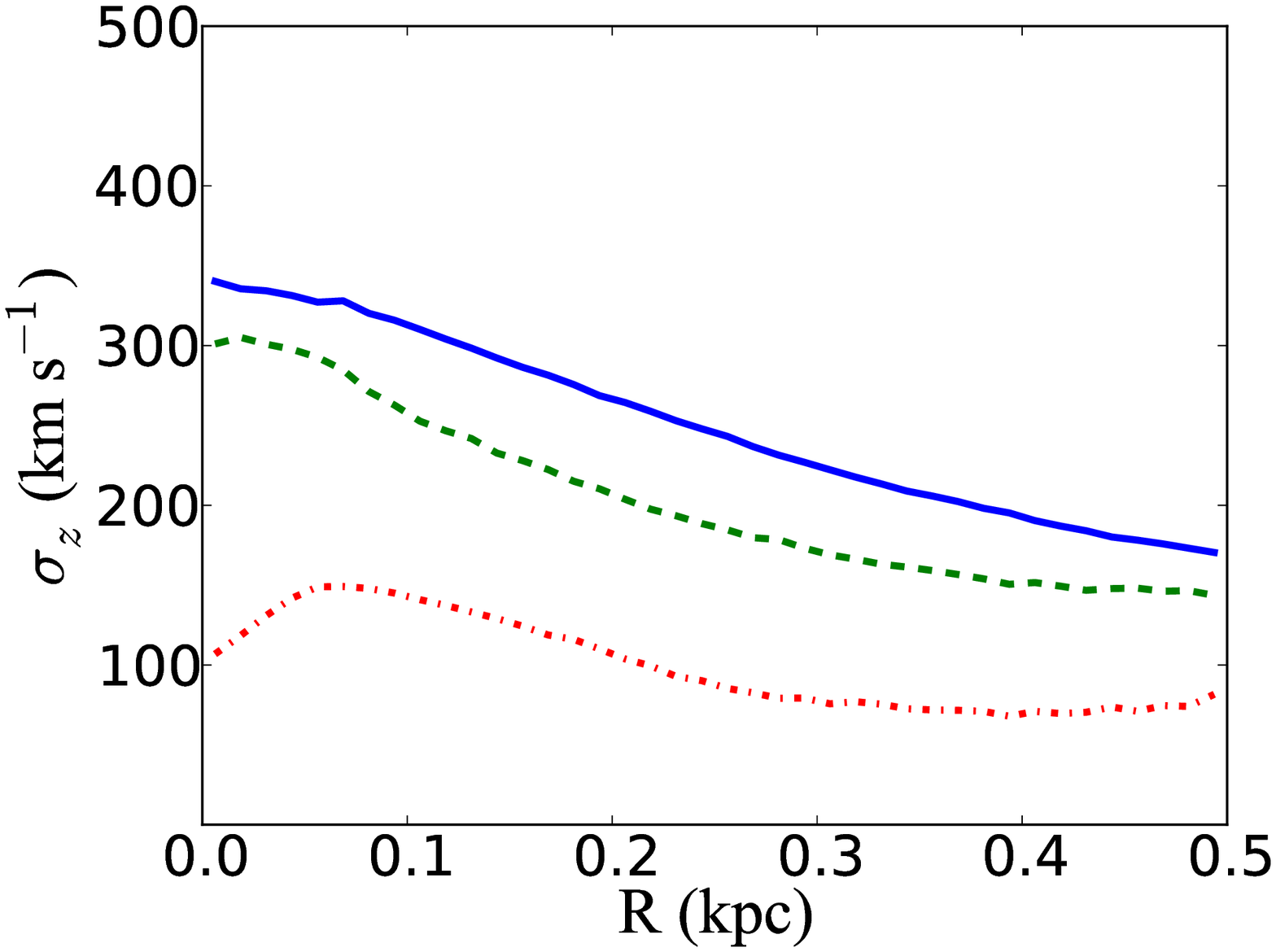}
\includegraphics[width=0.33\hsize,angle=0]{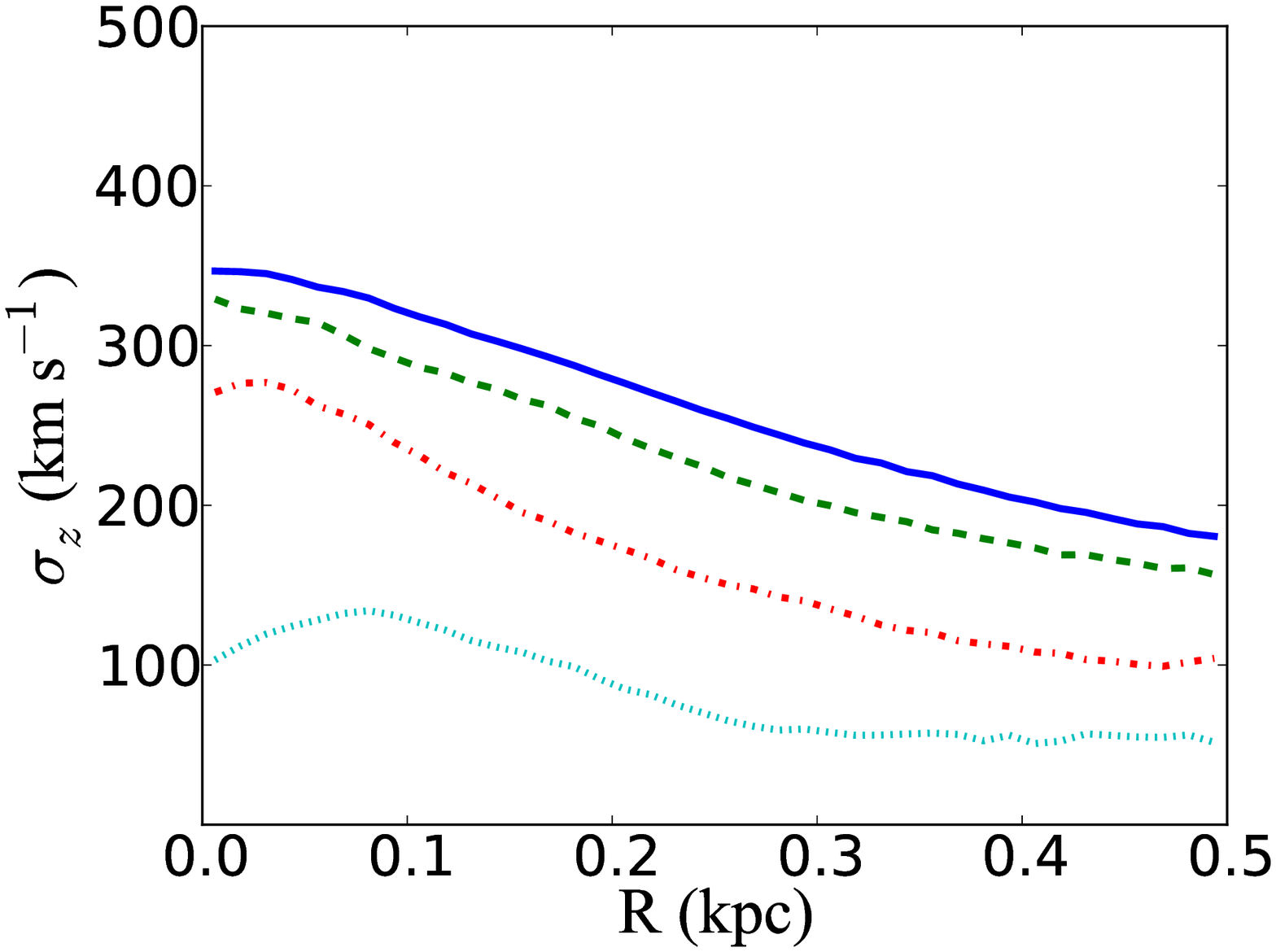}
\includegraphics[width=0.33\hsize,angle=0]{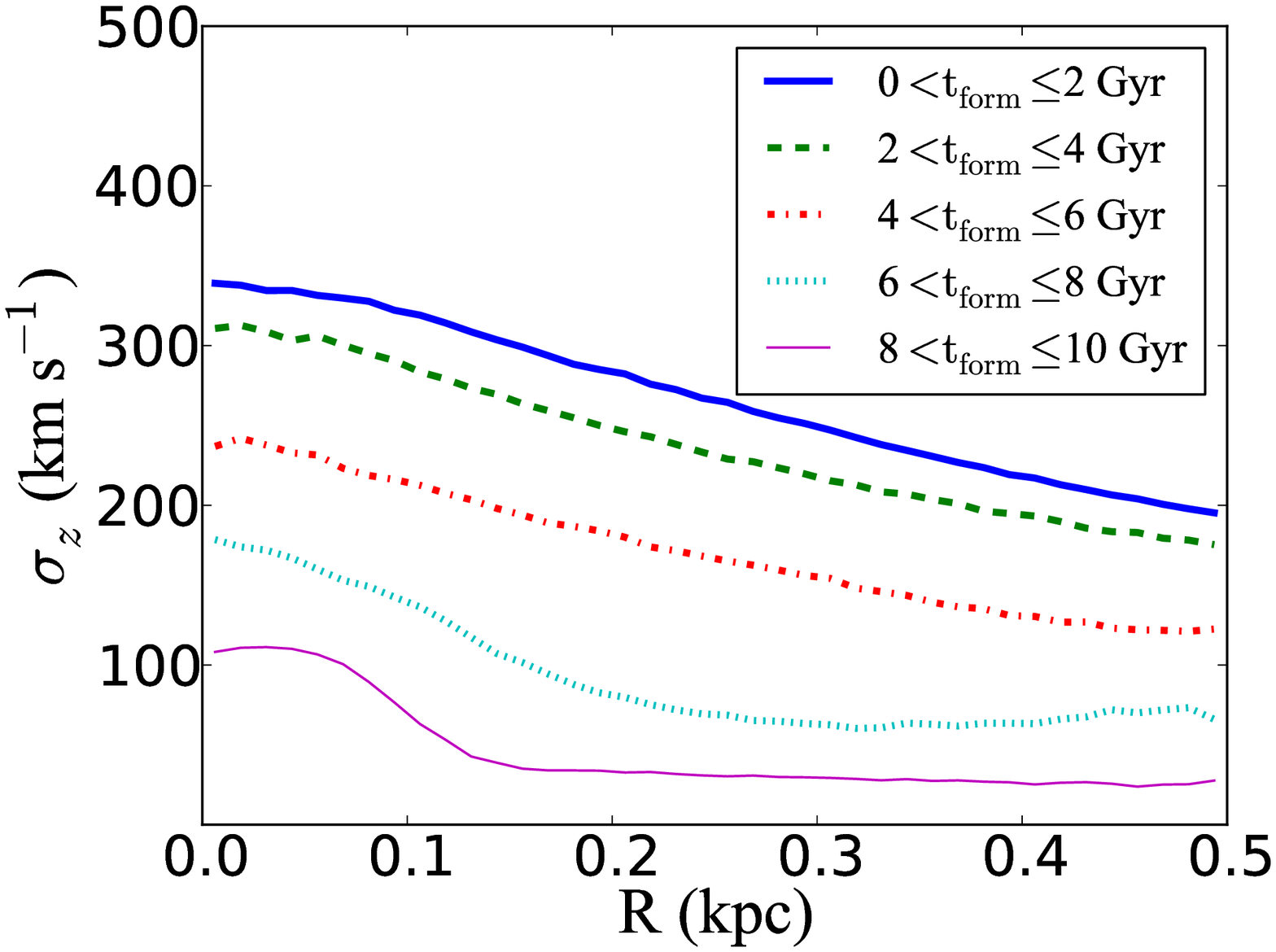}\\
  
\includegraphics[width=0.33\hsize,angle=0]{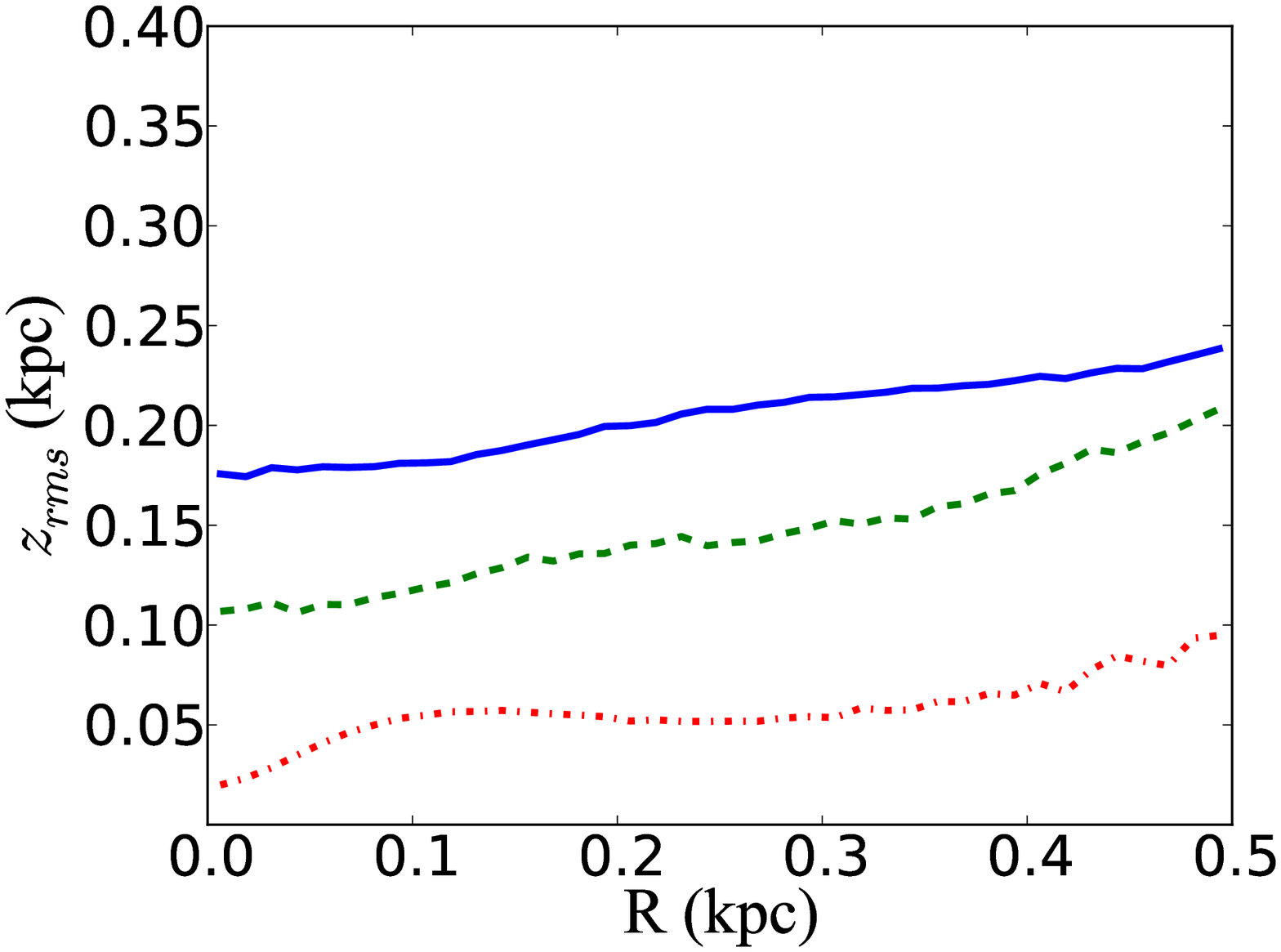}
\includegraphics[width=0.33\hsize,angle=0]{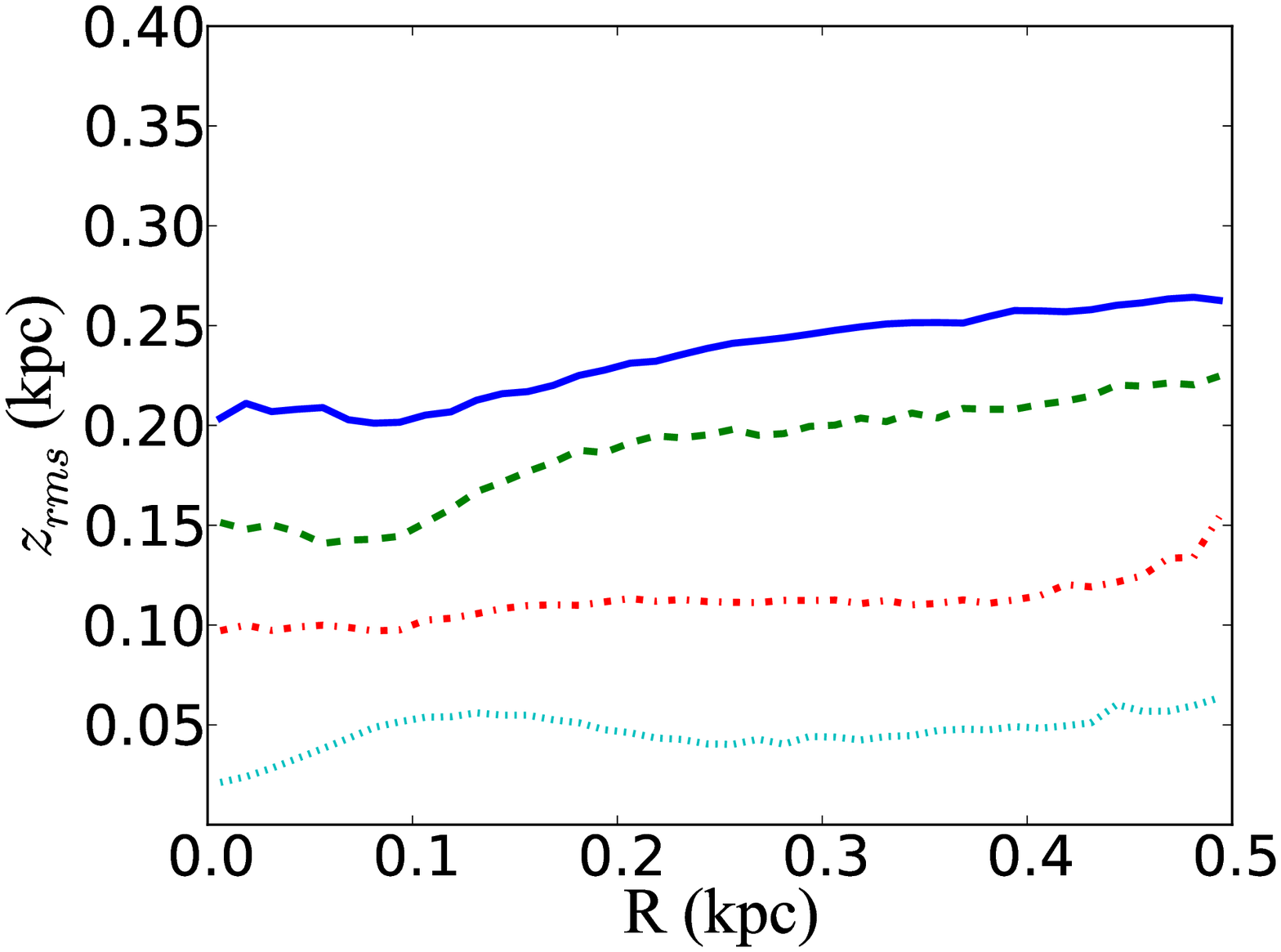}
\includegraphics[width=0.33\hsize,angle=0]{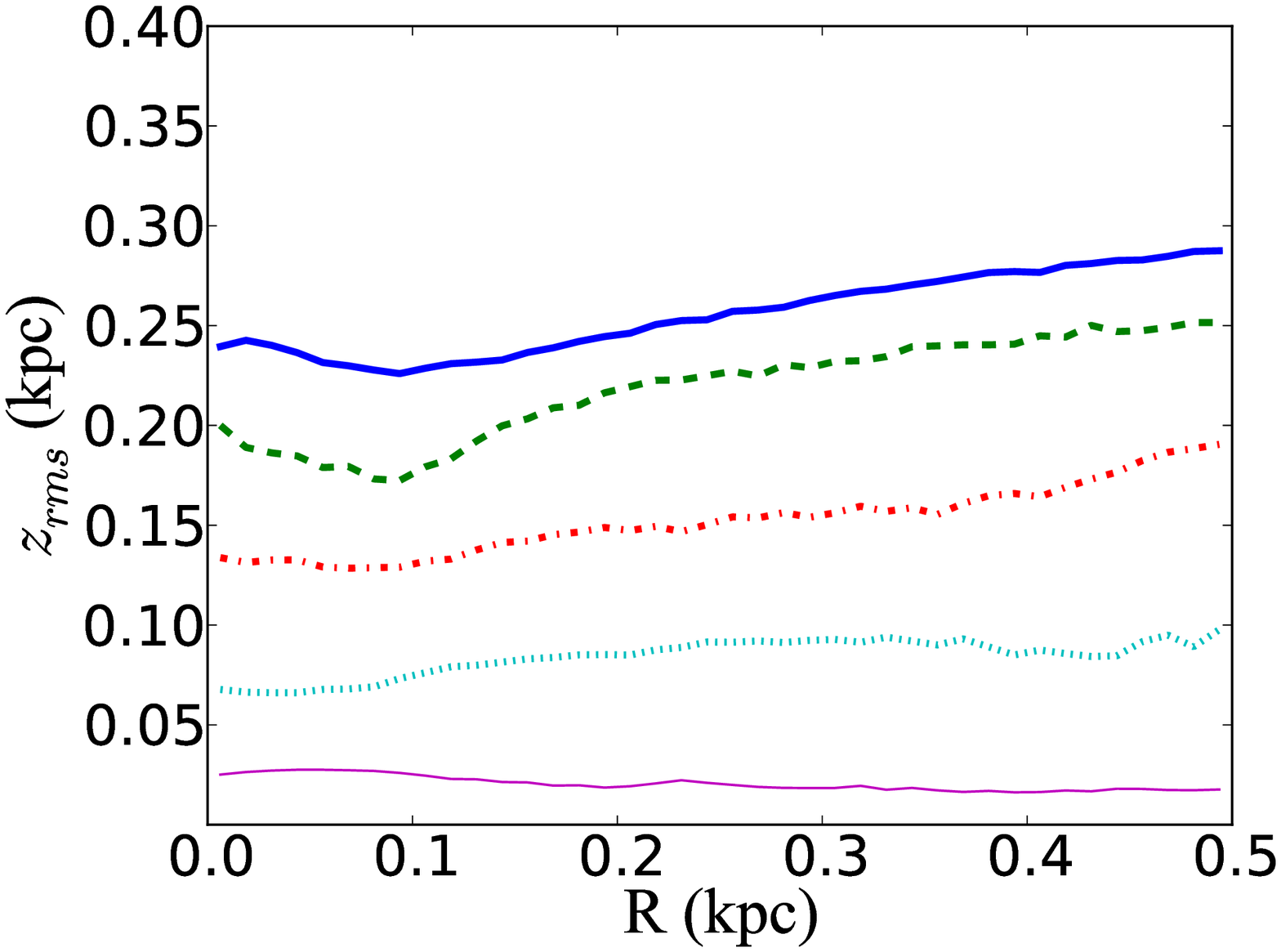}
 
\end{tabular}
\caption{Radial profile of the vertical velocity dispersion $\sigma_z$ (top panels),
  and $rms$ height  $z_{rms}$ (bottom panels),
  measured from the intrinsic mass-weighted kinematics and density distribution
  as a function of the age of the star particles after 6 (left-hand
  panels), 8 (central panels), and 10 Gyr (right-hand
  panels).  The coloured lines correspond to the
  profiles for stars formed at different times as indicated in the top
  right corner of the right-hand panel. The same colour marks the
  vertical velocity dispersion of the same stellar population evolving
  with time.}
\label{fig:sigmaz_evolution}
\end{figure*}

\subsection{Vertical velocity dispersion from the intrinsic mass-weighted kinematics}
\label{sec:intrinsic}

We calculate the vertical velocity dispersion of all the star
particles of the simulation with a formation time $t \geq t_{\rm in}$
from the intrinsic mass-weighted kinematics as
\begin{equation}
 \sigma_{z,t_{\rm in}}^2 = \frac{\displaystyle\sum_{i=1}^{N_{t_{\rm in}}} m_i v_{z,i}^2}{\displaystyle\sum_{i=1}^{N_{t_{\rm in}}} m_i}
\end{equation}
where $N_{t_{\rm in}}$ is the number of star particles in the given time range, while
$m_i$ and $v_{z,i}$ are the mass and vertical velocity of $i$-th
star particle, respectively.

When we consider all the simulation star particles at 10 Gyr, $\sigma_z$ steadily
increases from about 120 \kms\ at 1 kpc from the centre to a maximum
of about 270 \kms\ at about 0.1 kpc. At smaller radii, $\sigma_z$ drops
to a central minimum value of about 220
\kms\ (Figure~\ref{fig:sigmaz_age}).
Since these are the intrinsic kinematics, the \sd\ is not an artefact due to the method we adopted in
recovering the stellar kinematics. Moreover, the central dip in
$\sigma_z$ progressively fills in when removing star particles younger than
$0.2, 0.4, \ldots, 1.0, 1.5$ Gyr. The radial profile of $\sigma_z$
shows a central peak when all the stars younger than 2 Gyr are
removed (Figure~\ref{fig:sigmaz_age}). Therefore, we infer that the
central dip in $\sigma_z$, which we show below, and observe as a \sd\ in the recovered
luminosity-weighted kinematics, is driven by the star formation
of the galaxy centre.

To further investigate which stars contribute most to the \sd,
we plot in Figure~\ref{fig:sigmaz_evolution} the inner radial
profile of $\sigma_z$ and height $z$ for stars of different age bins when the
simulated galaxy is 6, 8 and 10 Gyr old.
At 6 Gyr, the stars younger than 2 Gyr show an almost constant
$\sigma_z$ between about 0.3 and 0.5 kpc. Inwards it rises to a
maximum of about 150 at about 0.1 kpc and drops to a central local
minimum of 100 \kms. The $\sigma_z$ radial profile of the stars with
an age between 2 and 4 Gyr and between 4 and 6 Gyr increases from
about 150 \kms\ at 0.5 kpc to about 300 and 350 \kms\ in the centre,
respectively. We interpret the larger $\sigma_z$ measured for the
older stellar populations due to the vertical heating of the stars,
which also results in a notable increase of their scale
height with time.
The same trend is observed at 8 and 10
Gyr: younger stars are characterized by a remarkably lower $\sigma_z$
with respect to the older ones and their $\sigma_z$ radial profile has
a pronounced central dip whereas older stars show a centrally peaked
$\sigma_z$. The vertical heating of the stellar orbits is more
effective in the first few Gyrs after formation. Afterwards both the shape and
amplitude of the $\sigma_z$ radial profile changes slowly with
time. These findings confirm that younger stars play a crucial role in
shaping the intrinsic mass-weighted kinematics in the centres of
galaxies. We expect to observe an even stronger effect on the
recovered luminosity-weighted kinematics since the younger stars are
much brighter than the older ones.
The vertical heating of the stellar orbits is also clear when we analyse the height, $z_{rms}$,
which increases as a function of time, as expected.
Therefore the stars first settle into a thin disc,
which becomes progressively thicker with time.

\begin{figure}
\centering
\includegraphics[width=0.95\hsize,angle=0]{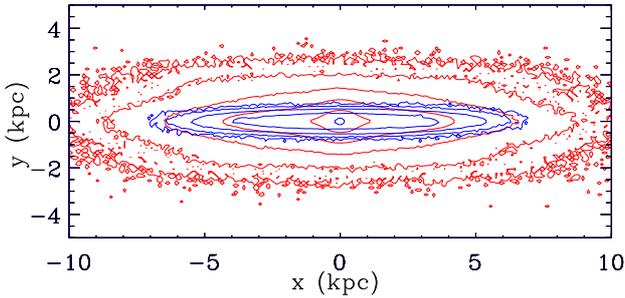}
\caption{Density contours of the galaxy at 10 Gyr as seen edge-on.
  The blue contours identify the distribution of young stars (age $\le 2$ Gyr),
  which is responsible for the \sd, while the red ones are for older star particles.
   A colour version of this figure is available online.}
\label{fig:spatial_distribution}
\end{figure}

Figure~\ref{fig:spatial_distribution} shows an edge-on view
  of the spatial distribution of the simulated galaxy at 10 Gyr.
  The blue contours identify the population with $t \le 2$ Gyr
  that is responsible for the \sd.

\subsection{LOS kinematic parameters from
  the recovered luminosity-weighted kinematics }
\label{sec:recovered}

\begin{figure*}
\centering
\begin{tabular}{c}
\includegraphics[width=0.96\hsize,angle=0]{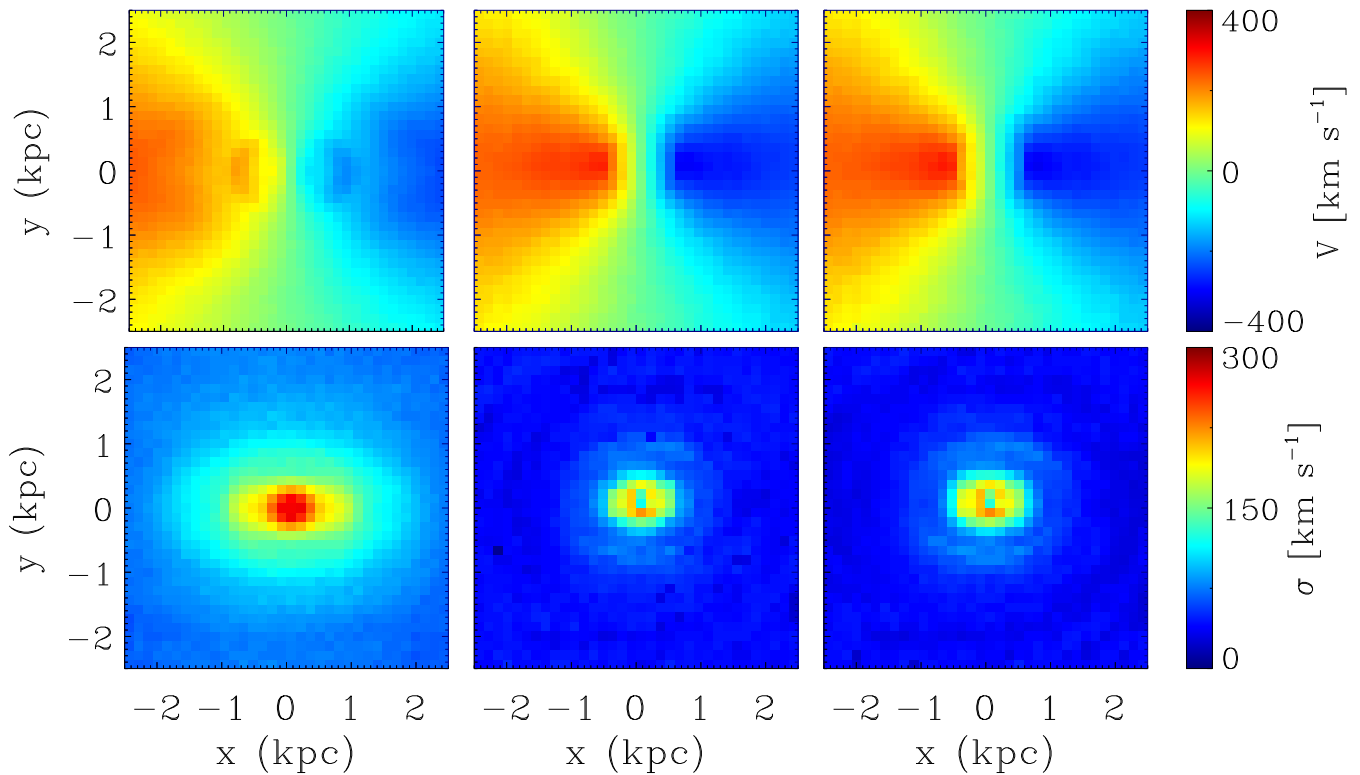}
\end{tabular}
\caption{Intrinsic mass-weighted kinematics (left-hand panels) and
  recovered luminosity-weighted kinematics measured in the spectral
  range covering the \MgI\ (central panels) and \CaII\ triplets
  (right-hand panels) for the stars of the galaxy at 10 Gyr
  seen at an inclination of $45^\circ$.  The LOS velocity (top panels) and velocity
  dispersion (bottom panels) are shown with the corresponding colour code at right.}
\label{fig:maps_45deg10Gyr}
\end{figure*}

In Figure~\ref{fig:maps_45deg10Gyr} we compare the maps of the intrinsic mass-weighted
kinematics (left-hand panels) to those measured from the spectral regions of the \MgI\  (middle panels) and
\CaII\ (right-hand panels) triplets (luminosity-weighted kinematics). The
galaxy is 10 Gyr old and is seen at an inclination $i=45^\circ$.  As anticipated, we find a pronounced \sd\ in the recovered
kinematics which is even deeper than that in the intrinsic kinematics.

\begin{figure}
\centering
\includegraphics[width=\hsize,angle=0]{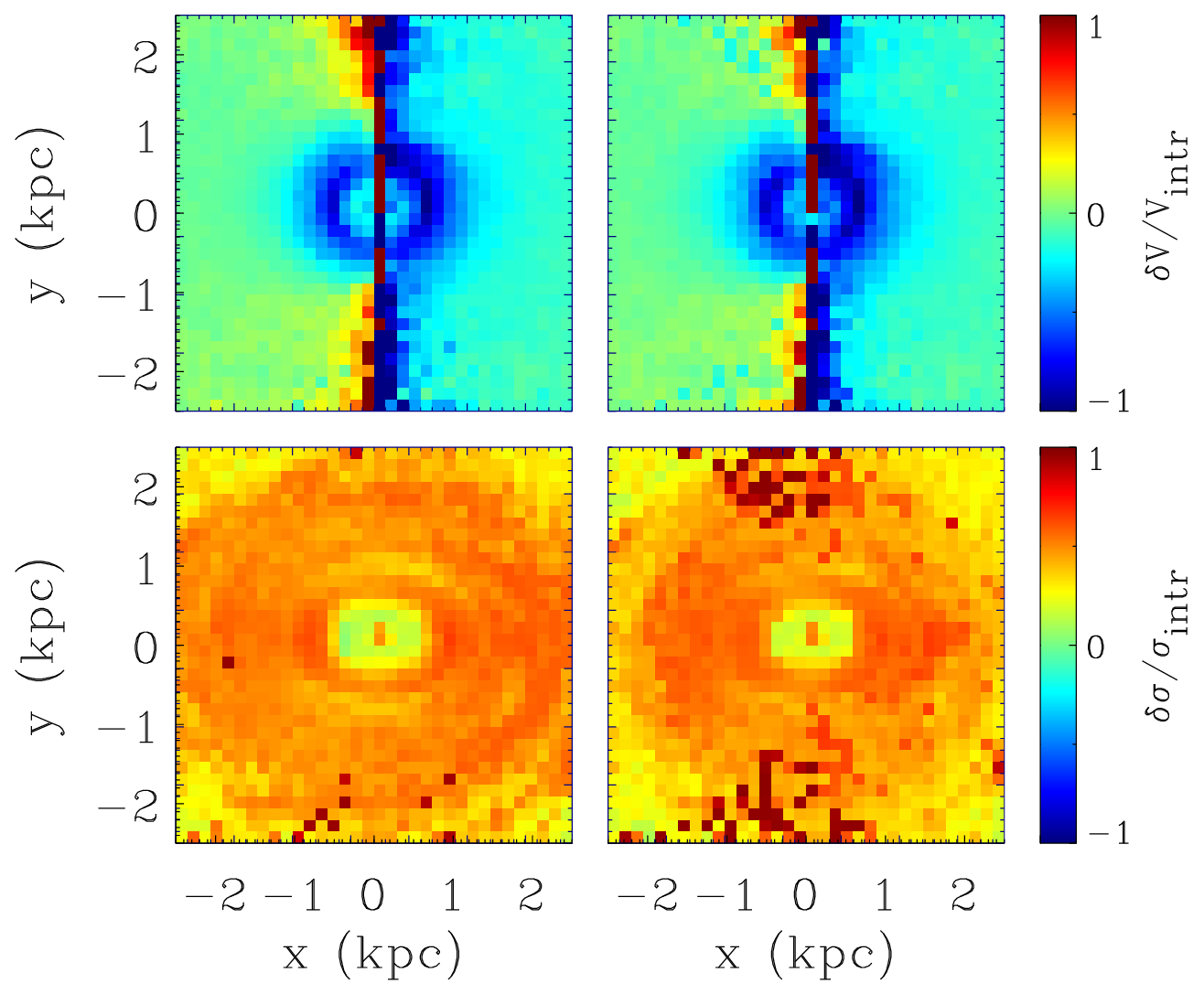}
\caption{Maps of the residuals of the recovered
  luminosity-weighted kinematics measured in the spectral range
  covering the \MgI\ (left-hand panels) and \CaII\ triplets
  (right-hand panels) with respect to the intrinsic mass-weighted
  kinematics for the stars of the simulated galaxy at 10 Gyr seen at
  an inclination of $45^\circ$. The residuals of the LOS velocity (top panels) and
  velocity dispersion (bottom panels) are shown
  with the corresponding colour code at the right.
  The size and orientation of the field of view are the same as
  in Figure~\ref{fig:maps_45deg10Gyr}.}
\label{fig:residuals_45deg10Gyr}
\end{figure}

\begin{figure*}
\centering
\includegraphics[width=0.99\hsize]{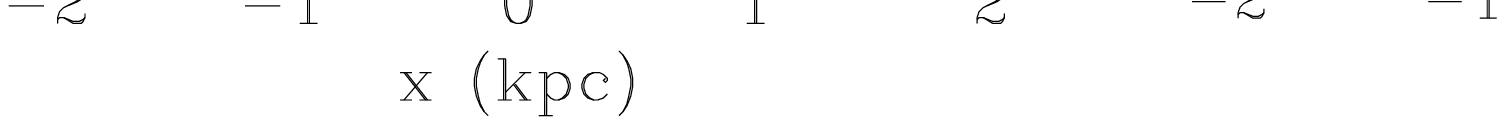}
\caption{Intrinsic mass-weighted kinematics (solid black lines) and
  recovered luminosity-weighted kinematics measured in the spectral
  ranges covering the \MgI\ (green circles) and \CaII\ triplets (red
  triangles) along the major axis of the simulated galaxy at 5 Gyr (left-hand panels)
  and 10 Gyr (right-hand panels) seen
  at an inclination of $45^\circ$. The radial profiles of the LOS
  velocity are shown at top and velocity dispersion are shown at bottom.}
\label{fig:majoraxis_45deg5+10Gyr}
\end{figure*}

In Figure~\ref{fig:residuals_45deg10Gyr} we plot the residual maps
we obtained by subtracting the recovered luminosity-weighted
kinematics from the intrinsic mass-weighted one and dividing the
difference by the intrinsic one to highlight the differences in the
LOS kinematic parameters due to the adopted spectral range.
The strongest fractional velocity residuals are along the minor-axis direction, when the intrinsic value is close to 0 \kms.
We find a good agreement in the velocity and velocity dispersion maps:
the differences between the intrinsic and the recovered kinematics are compatible with the error bars,
as visible in Figure~\ref{fig:majoraxis_45deg5+10Gyr} (right-hand panel).

We extract the radial profiles of LOS kinematic parameters from
both the recovered and intrinsic kinematics in a 0.15-kpc wide strip
along the major axis of the galaxy at 10 Gyr ($i=45^\circ$) and plot them in Figure~\ref{fig:majoraxis_45deg5+10Gyr}.
We decrease the bin sizes of the central regions to highlight the kinematics in that area. 
The intrinsic deprojected rotation curve rises to a maximum of about 220 km
s$^{-1}$ at about 1 kpc from the centre and remains almost flat
outwards, whereas the recovered rotation velocity is characterized by
a peak velocity of about 300 \kms\ at about 0.5 kpc and gently
decreases to match the intrinsic velocity at about 2.5 kpc.
The intrinsic velocity dispersion shows a central local minimum at
about 220 \kms. It rises to about 270 \kms\ at about 0.2 kpc and then
it decreases almost exponentially outwards. In the inner $0.5$ kpc
 the recovered velocity dispersion flattens at about
200 \kms\ with a central sharp drop to about 100 \kms . The radial
profiles of the intrinsic and recovered velocity dispersion profiles
are very different also in the outer regions where the recovered
velocity dispersion shows a marked decline to about 50 \kms\ at about
1 kpc and flattens out at larger radii.
The same LOS kinematic features are observed along the galaxy major
axis when it is 5 Gyr old
(Figure~\ref{fig:majoraxis_45deg5+10Gyr}). Indeed the intrinsic
mass-weighted and recovered luminosity-weighted kinematics are
remarkably similar to those extracted from the 10 Gyr simulation
time-step. In particular, the \sd\ is clearly visible and more
pronounced in the velocity dispersion profile obtained from the
recovered luminosity-weighted kinematics.


\begin{figure*}
\centering
\begin{tabular}{c}
  \includegraphics[width=0.98\hsize,angle=0]{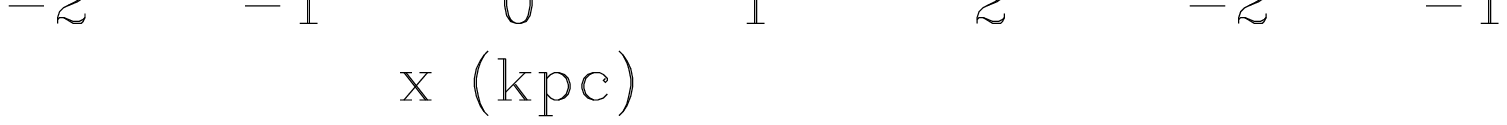} 
\end{tabular}
\caption{The radial profiles of the LOS velocity and velocity dispersion
  (from top to bottom) along the major axis of the simulated galaxy at 10 Gyr seen at an
  inclination of $45^\circ$ for stars of different age (solid colour
  lines coded as indicated in the left-hand panel) from the
  kinematics recovered in the spectral ranges covering the
  \MgI\ triplet lines (left-hand panels) and \CaII\ triplet (right-hand
  panels).  The solid and dashed
  black lines correspond to the recovered luminosity-weighted and
  intrinsic mass-weighted kinematics obtained including all the
  simulation star particles, respectively.}
\label{fig:majoraxis_45deg10Gyr_agebins}
\end{figure*}

To investigate the possibility that the recovered kinematics at the
galaxy centre are biased to higher rotation velocities and lower
velocity dispersions due to the presence of young stars in a
thin and kinematically-cool nuclear disc, we show in
Figure~\ref{fig:majoraxis_45deg10Gyr_agebins} the LOS kinematics
of stars of different ages.
This analysis demonstrates that young stars exhibit
  a significantly lower velocity dispersion and their rotation velocity is systematically higher
  than the oldest counterpart, which instead shows a peaked distribution of the velocity dispersion.
  Furthermore, there are no significant differences between the kinematics recovered in the two
 spectral ranges of \MgI\ and \CaII\ triplets for the stars in the same
age bin.
%

A centrally-peaked profile is measured in the velocity dispersion of
all the age bins. The central value increases from $\approx$ 100 \kms\ for
the stars younger than 2 Gyr, to $\approx$ 200 \kms\ for the $2-4$ Gyr-old
stars and to $\approx$ 300 \kms\ for the stars older than 4
Gyr. The velocity dispersion profile is quite flat in the centre for
stars younger than 2 Gyr and it shows two symmetric local maxima or
bumps for stars with age between $0-$ and $6$ and $6-$ and $10$ Gyr, respectively. They correspond
to the local minima observed in the rotation velocity curves, which
all follow the same double-hump profile. This is characterized by an
inner steep slope of the rotation velocity, reaching a local
maximum followed by a slight drop to a local minimum. At larger radii
the rotation velocity of all the age bins rises again, reaches an
absolute maximum and then declines smoothly to match the intrinsic
rotation velocity at 2.5 kpc. The recovered rotation curve
from all the stars of the galaxy is rising out almost linearly
to about 0.6 kpc with a small change in slope seen
in the kinematics recovered in the \MgI\ spectral range. We conclude
that both the shoulder of the rotation curve and the central drop of
the velocity dispersion profile measured in the recovered kinematics
are due to the younger stars.


\section{Discussion and conclusions}
\label{sec:conclusions}

We have studied in detail the stellar kinematics of an unbarred galaxy with a
high-resolution $N$-body $+$ SPH simulation.
The intrinsic kinematics are obtained by weighting the velocity of the
star particles with their mass and are compared with the
luminosity-weighted ones that we recover from synthetic spectra built
by using our code SYNTRA, which combines the outputs of
the simulation at different time-steps with stellar population
synthesis models by \citet{Bruzual2003} to generate close-to-real
spectroscopic data. We measured
the luminosity-weighted stellar kinematics in two wavelength ranges
covering the \MgI\ triplet at $\lambda\ 5164, 5173, 5184$ \AA\ and the
\CaII\ triplet at $\lambda\ 8498, 8542, 8662$ \AA,
without finding any substantial difference in the LOS velocity and
velocity dispersion maps.

We notice remarkable differences in the rotation velocity curves and
velocity dispersion profiles of the recovered luminosity-weighted and
intrinsic mass-weighted kinematics. We demonstrate
that observational data are biased by the presence of young stars
in a nuclear disc. In our simulation such a disc is thin and
kinematically cool and the complex kinematics in the central regions
are a result of the superposition of older and younger stars.
This should not be confused with the nuclear stellar discs, which are found
in the centre of many galaxies, with scale-lengths $\approx 10-50$ pc
and luminosities $L_V \approx 10^6-10^7$ L$_{\odot}$ \citep{Pizzella2002,Morelli2010,Corsini2016},
which are the smallest and
brightest stellar discs known to date when compared to the embedded
discs of early-type galaxies ($h \approx 100$ pc) and to the main discs of lenticular and
spiral galaxies \citep{vandenBosch1998}.
The young disc found in the simulation can be better associated with a pseudobulge,
because it is characterized by different coexisting components \citep{Fabricius2012, MendezAbreu2014,deLorenzo2012}.
The total mass of the young stars ($t \le 2$ Gyr) represents
10\% of the total mass of all the star particles,
while it contributes up to 50\% to the total luminosity.
The SFR of the young population is 1.1 M$_{\odot}$ yr$^{-1}$,
in good agreement with the findings of other studies \citep{Wozniak2003, Wozniak2006}.

We find a significant drop in the velocity dispersion profile in
our recovered luminosity-weighted kinematics, much more than that in
the intrinsic mass-weighted one. In the simulation
the inflow of gas to the central region of the galaxy creates a thin
kinematically cool disc where stars form, reducing the central
velocity dispersion. The fact that the \sd\ is more pronounced in the
luminosity-weighted kinematic maps is a clear sign that the younger
stars dominate the light distribution in the centre,
although their contribution to the galaxy mass is less
significant. This result demonstrates that a
younger, luminous but not massive stellar population can
bias the recovered gravitational potential of galaxies when
weighted by luminosity. The visibility of \sd s strongly depends on the
mass ratio between the old dynamically-hot population and the young
 dynamically-cool one. This finding is extremely important to
correctly interpret the kinematics of the
central regions of galaxies, such as for the derivation of black
hole masses using stellar kinematics. In our case, the \sd\ is
linked to the presence of a central disc, as is clear from the
double-hump shape of the rotation velocity curve. 

Integral field spectroscopy of nearby galaxies can probe the full
extent and amplitude of \sd s and to which stellar component they
are associated, providing information on the current stellar
populations and their star formation history. In this sense, \sd s
can shed light on to the current dynamical state of galaxy central
regions. In this context, we highlight that our code SYNTRA represents an extremely useful
tool for building mock observations that can be used to recover
luminosity-weighted kinematics and analyse
simulations that include multiple stellar populations, for
example kinematic cores and counter-rotating discs, but also galaxies with both
thick and thin disc components. We would be able to disentangle the
contribution of these components and compare simulation results with
those found in recent integral-field spectroscopic surveys, such as
ATLAS3D \citep{Cappellari2011} and CALIFA \citep{Marmol2011}.

\section{Acknowledgements}

EP acknowledges the Jeremiah Horrocks Institute of the University of
Central Lancashire for the hospitality while this paper was in
progress. EP was partially supported by Accademia Nazionale dei Lincei
and Fondazione Ing. Aldo Gini.  Additional support while the paper was
completed was provided by the Italian Ministry for Education
University and Research (MIUR).  VPD was supported
by STFC Consolidated grants no, ST/J001341/1 and no. ST/M000877/1.
VPD acknowledges being a part of the network supported by
the COST Action TD1403 "Big Data Era in Sky and Earth Observation".
DRC was supported by STFC Consolidated grant no. ST/J001341/1.  EMC, EDB,
LM, and AP are supported by Padua University through grants
60A02-5857/13, 60A02-5833/14, and 60A02-4434/15 and BIRD 164402/16.
The simulation was run at the DiRAC Shared Memory Processing
system at the University of Cambridge, operated by the COSMOS Project
at the Department of Applied Mathematics and Theoretical Physics on
behalf of the STFC DiRAC HPC Facility (www.dirac.ac.uk). This
equipment was funded by BIS National E-infrastructure capital grant
ST/J005673/1, STFC capital grant ST/H008586/1, and STFC DiRAC
Operations grant ST/K00333X/1. DiRAC is part of the National
E-Infrastructure. EP would like to thank R. Ragazzoni
for his support while this work was in progress.

\bibliographystyle{mn2e1}
\bibliography{708main}{}

\begin{thebibliography}{}

\bibitem[\protect\citeauthoryear{Athanassoula \& Misiriotis}{2002}]{Athanassoula2002} 
Athanassoula E., Misiriotis A., 2002, MNRAS, 330, 35 

\bibitem[\protect\citeauthoryear{Bender}{1990}]{Bender1990} 
Bender R., 1990, A\&A, 229, 441 

\bibitem[\protect\citeauthoryear{Bertelli et al.}{1994}]{Bertelli1994}
  Bertelli G., Bressan A., Chiosi C., Fagotto F., Nasi E., 1994, A\&AS, 106,  

\bibitem[\protect\citeauthoryear{Bertola et al.}{1996}]{Bertola1996}
  Bertola F., Cinzano P., Corsini E.~M., Pizzella A., Persic M., Salucci P., 1996, ApJ, 458, L67 

\bibitem[\protect\citeauthoryear{Binney}{1980}]{Binney1980} 
Binney J., 1980, MNRAS, 190, 873

\bibitem[\protect\citeauthoryear{Bottema}{1989}]{Bottema1989} 
Bottema R., 1989, A\&A, 221, 236 

\bibitem[\protect\citeauthoryear{Bottema \& Gerritsen}{1997}]{Bottema1997} 
Bottema R., Gerritsen J.~P.~E., 1997, MNRAS, 290, 585 

\bibitem[\protect\citeauthoryear{Bruzual \& Charlot}{2003}]{Bruzual2003} 
Bruzual G., Charlot S., 2003, MNRAS, 344, 1000 

\bibitem[\protect\citeauthoryear{Bureau \& Athanassoula}{2005}]{Bureau2005} 
Bureau M., Athanassoula E., 2005, ApJ, 626, 159 

\bibitem[\protect\citeauthoryear{Cappellari \& Emsellem}{2004}]{Cappellari2004} 
Cappellari M., Emsellem E., 2004, PASP, 116, 138 

\bibitem[\protect\citeauthoryear{Cappellari et al.}{2011}]{Cappellari2011}
Cappellari M., et al., 2011, MNRAS, 413, 813 

\bibitem[\protect\citeauthoryear{Chung \& Bureau}{2004}]{Chung2004}
  Chung A., Bureau M., 2004, AJ, 127, 3192 

\bibitem[\protect\citeauthoryear{Ciotti \& Lanzoni}{1997}]{Ciotti1997}
  Ciotti L., Lanzoni B., 1997, A\&A, 321, 724 

\bibitem[\protect\citeauthoryear{Coccato et al.}{2015}]{Coccato2015}
  Coccato L., et al., 2015, A\&A, 581, A65 

\bibitem[\protect\citeauthoryear{Cole et al.}{2014}]{Cole2014} 
Cole D.~R., Debattista V.~P., Erwin P., Earp S.~W.~F., Ro{\v s}kar R., 2014, MNRAS, 445, 3352 

\bibitem[\protect\citeauthoryear{Comer{\'o}n et al.}{2008}]{Comeron2008} 
Comer{\'o}n S., Knapen J.~H., Beckman J.~E., 2008, A\&A, 485, 695 

\bibitem[\protect\citeauthoryear{Corsini et al.}{2016}]{Corsini2016}
  Corsini E.~M., Morelli L., Pastorello N., Dalla Bont{\`a} E., Pizzella A., Portaluri E., 2016, MNRAS, 457, 1198 

\bibitem[\protect\citeauthoryear{de Lorenzo-C{\'a}ceres et al.}{2012}]{deLorenzo2012} 
de Lorenzo-C{\'a}ceres A., Vazdekis A., Aguerri J.~A.~L., Corsini E.~M., Debattista V.~P., 2012, MNRAS, 420, 1092 

\bibitem[\protect\citeauthoryear{Dressler \& Richstone}{1990}]{Dressler1990} 
Dressler A., Richstone D.~O., 1990, ApJ, 348, 120 

\bibitem[\protect\citeauthoryear{Emsellem et al.}{2001}]{Emsellem2001} 
Emsellem E., Greusard D., Combes F., Friedli D., Leon S., P{\'e}contal E., Wozniak H., 2001, A\&A, 368, 52 
\bibitem[\protect\citeauthoryear{Erwin et al.}{2015}]{Erwin2015}
  Erwin P., et al., 2015, MNRAS, 446, 4039 

\bibitem[\protect\citeauthoryear{Fabricius et al.}{2012}]{Fabricius2012} 
Fabricius M.~H., Saglia R.~P., Fisher D.~B., Drory N., Bender R., Hopp U., 2012, ApJ, 754, 67 

\bibitem[\protect\citeauthoryear{Fabricius et al.}{2014}]{Fabricius2014} Fabricius M.~H., et al., 2014, MNRAS, 441, 2212 

\bibitem[\protect\citeauthoryear{Falc{\'o}n-Barroso et al.}{2006}]{FalconBarroso2006}
Falc{\'o}n-Barroso J., et al., 2006, MNRAS, 369, 529 

\bibitem[\protect\citeauthoryear{Friedli}{1996}]{Friedli1996} 
Friedli D., 1996, A\&A, 312, 761 

\bibitem[\protect\citeauthoryear{Gerhard}{1993}]{Gerhard1993} 
Gerhard O.~E, 1993, MNRAS, 265, 213 

\bibitem[\protect\citeauthoryear{Graham et al.}{1998}]{Graham1998} 
Graham A.~W., Colless M.~M., Busarello G., Zaggia S., Longo G., 1998, A\&AS, 133, 325 

\bibitem[\protect\citeauthoryear{Jarvis et al.}{1988}]{Jarvis1988}
  Jarvis B.~J., Dubath P., Martinet L., Bacon R., 1988, A\&AS, 74, 513 

\bibitem[\protect\citeauthoryear{Koleva et al.}{2008}]{Koleva2008} 
Koleva M., Prugniel P., De Rijcke S., 2008, AN, 329, 968 

\bibitem[\protect\citeauthoryear{Krajnovi{\'c} et al.}{2008}]{Krajnovic2008}
Krajnovi{\'c} D., et al., 2008, MNRAS, 390, 93 

\bibitem[\protect\citeauthoryear{Krajnovi{\'c} et al.}{2013}]{Krajnovic2013}
  Krajnovi{\'c} D., et al., 2013, MNRAS, 433, 2812

\bibitem[\protect\citeauthoryear{Loebman et al.}{2011}]{Loebman2011}
  Loebman S.~R., Ro{\v s}kar R., Debattista V.~P., Ivezi{\'c} {\v Z}., Quinn T.~R., Wadsley J., 2011, ApJ, 737, 8 

\bibitem[\protect\citeauthoryear{M{\'a}rmol-Queralt{\'o} et al.}{2011}]{Marmol2011}
  M{\'a}rmol-Queralt{\'o} E., et al., 2011, A\&A, 534, A8 

\bibitem[\protect\citeauthoryear{M{\'e}ndez-Abreu et al.}{2014}]{MendezAbreu2014} 
M{\'e}ndez-Abreu J., Debattista V.~P., Corsini E.~M., Aguerri J.~A.~L., 2014, A\&A, 572, 25

\bibitem[\protect\citeauthoryear{Miller \& Scalo}{1979}]{Miller1979}
  Miller G.~E., Scalo J.~M., 1979, ApJS, 41, 513

\bibitem[\protect\citeauthoryear{Morelli et al.}{2010}]{Morelli2010}
  Morelli L., Cesetti M., Corsini E.~M., Pizzella A., Dalla Bont{\`a} E., Sarzi M., Bertola F., 2010, A\&A, 518, A32 

\bibitem[\protect\citeauthoryear{Pizzella et al.}{2002}]{Pizzella2002}
  Pizzella A., Corsini E.~M., Morelli L., Sarzi M., Scarlata C., Stiavelli M., Bertola F., 2002, ApJ, 573, 131 

\bibitem[\protect\citeauthoryear{P{\'e}rez, S{\'a}nchez-Bl{\'a}zquez, \& Zurita}{2009}]{Perez2009} 
P{\'e}rez I., S{\'a}nchez-Bl{\'a}zquez P., Zurita A., 2009, A\&A, 495, 775 

\bibitem[\protect\citeauthoryear{Raiteri, Villata, \& Navarro}{1996}]{Raiteri1996}
  Raiteri C.~M., Villata M., Navarro J.~F., 1996, A\&A, 315, 105 

\bibitem[\protect\citeauthoryear{Salpeter}{1995}]{Salpeter1955} 
Salpeter, E. E., 1955, ApJ, 121, 161

\bibitem[\protect\citeauthoryear{Shen et al.}{2010}]{Shen2010}
  Shen J., Rich R.~M., Kormendy J., Howard C.~D., De Propris R., Kunder A., 2010, ApJ, 720, L72 

\bibitem[\protect\citeauthoryear{Stinson et al.}{2006}]{Stinson2006}
  Stinson G., Seth A., Katz N., Wadsley J., Governato F., Quinn T., 2006, MNRAS, 373, 1074 

\bibitem[\protect\citeauthoryear{van den Bosch, Jaffe, \& van der Marel}{1998}]{vandenBosch1998}
  van den Bosch F.~C., Jaffe W., van der Marel R.~P., 1998, MNRAS, 293, 343 

\bibitem[\protect\citeauthoryear{van der Marel \& Franx}{1993}]{vanderMarel1993}
van der Marel R.~P., Franx M., 1993, ApJ, 407, 525 

\bibitem[\protect\citeauthoryear{van der Marel}{1994}]{vanderMarel1994} 
van der Marel R.~P., 1994, MNRAS, 270, 271 

\bibitem[\protect\citeauthoryear{Wadsley, Stadel, \& Quinn}{2004}]{Wadsley2004}
  Wadsley J.~W., Stadel J., Quinn T., 2004, NewA, 9, 137

\bibitem[\protect\citeauthoryear{Weidemann}{1987}]{Weidemann1987}
  Weidemann V., 1987, A\&A, 188, 74 

\bibitem[\protect\citeauthoryear{Woosley \& Weaver}{1995}]{Woosley1995}
  Woosley S.~E., Weaver T.~A., 1995, ApJS, 101, 181 

\bibitem[\protect\citeauthoryear{Wozniak et  al.}{2003}]{Wozniak2003} 
Wozniak H., Combes F., Emsellem E., Friedli D., 2003, A\&A, 409, 469 

\bibitem[\protect\citeauthoryear{Wozniak \& Champavert}{2006}]{Wozniak2006} 
Wozniak H., Champavert N., 2006, MNRAS, 369, 853 

\end{thebibliography}

\label{lastpage}
\end{document}